\documentclass[11pt, a4paper]{article}                                                        
\usepackage[utf8]{inputenc}
\usepackage[british]{babel}
\usepackage[nottoc]{tocbibind}
\usepackage{authblk}

\usepackage{blindtext}
\usepackage[top=1in, bottom=1in,left=1in,right=1in]{geometry}
\usepackage{amssymb}
\usepackage{amsmath}
\usepackage{mathtools}
\usepackage{multicol}
\usepackage{tikz}
\DeclareMathOperator{\sech}{sech}
\usepackage{amsfonts}
\usepackage{graphicx}
\usepackage{subfig}
\usepackage{caption}
\usepackage{braket}
\usepackage{array}
\usepackage{float}
\usepackage{hyperref}
\usepackage{cite}
\newcolumntype{M}[1]{>{\centering\arraybackslash}m{#1}}
\newcolumntype{N}{@{}m{0pt}@{}}
\usepackage{enumerate}
\usepackage{bm}
\usepackage{lipsum}

\providecommand{\keywords}[1]
{
  \small	
  \textbf{\textit{Keywords---}} #1
}

\begin{document}

\title{\textbf{{Atomic Inversion and Entanglement Dynamics in the Mixed Squeezed Coherent State Version of the Jaynes-Cummings Interaction}}}
\def\correspondingauthor{\footnote{Corresponding author's email: mandalkoushik1993@gmail.com}}
\author[1]{Koushik Mandal \correspondingauthor{}}
\author[]{Pooja Jethwani}
\author[]{ M. V. Satyanarayana }
\affil[]{\textit{Department of Physics, Indian Institute of Technology Madras, Chennai 600036, India}}
\date{}

\maketitle

\abstract{Coherent signal containing squeezed noise in a mixed state  of radiation field is considered here as a non-Gaussian mixture of a coherent state density operator and a squeezed state density operator, as opposed to the usual well known squeezed  coherent state. Both these states are `quantum' noise-included signal states. Effects of these two distinct ways of adding squeezing to a coherent state are compared and contrasted. The main objective of this work is to study the mixed state version of the Jaynes-Cummings model in the context of a two-level atom interacting with a mixed field state of a squeezed vacuum and a coherent state. The pure squeezed coherent state (PSCS) and the mixed squeezed coherent state (MSCS) are used as the states of the radiation field. The photon-counting distribution (PCD), the atomic inversion and the entanglement dynamics of atom-field interaction for both the radiation fields are investigated and compared with each other. We observe that depending on the state of the field, squeezing has very different effects on coherent photons. Mild squeezing on the coherent photons strongly localizes the PCD for PSCS; however, for MSCS there is no such localization observed - instead squeezing manifests for MSCS as oscillations in the PCD. The effects of squeezing on the atomic inversion and the entanglement dynamics for MSCS are contrasting in comparison with the corresponding quantities associated with PSCS. It is well known in the literature that for PSCS, increasing the squeezing increases the well-known ringing revivals in the atomic inversion, and also increases irregularity in the entanglement dynamics. However, increasing the squeezing in MSCS very significantly alters the collapse-revival pattern in the atomic inversion and the entanglement dynamics of the Jaynes-Cummings model. For MSCS, the effect of squeezing on Mandel's $Q$ parameter and Wigner functions are also presented. }\\


\section{Introduction}

The study of interactions between atoms and fields is one of the major topics in quantum optics. The quantities like the photon counting distribution, the atomic inversion, and the entanglement dynamics have been at the centre of interest in quantum optics. Also, entanglement is an important aspect of quantum information \cite{RevModPhys.81.865}, quantum cryptography \cite{PhysRevLett.67.661}, quantum teleportation \cite{PhysRevLett.70.1895}, super dense coding \cite{PhysRevLett.69.2881} etc. The interaction between the radiation field and the two-level atoms provides a way to study the entanglement dynamics between the two systems.

The use of density operators of the form
\begin{equation}
    \hat{\rho} = \sum_{n = 1}^{N} \rho_{nm} \ket{n}\bra{n} + \sum_{n = 1}^{N} \sum_{n'>n} (\rho_{nn'} \ket{n}\bra{n'} + \rho^{*}_{nn'} \ket{n'}\bra{n}),
\end{equation}
is well known in literature in the context of quantum information \cite{nielsen2002quantum, wiseman2009quantum, PhysRevA.63.044304}, quantum control of dissipative systems\cite{schirmer2002quantum, solomon2004dissipative}. It is to be noted that a convex combination of density operators can be cast in the above form. Recently, the study of entanglement for a mixed states are of high interest. In references \cite{PhysRevLett.101.260505, PhysRevLett.125.200501, PhysRevLett.128.140502, PhysRevLett.118.040801, PhysRevA.72.012315, PhysRevA.75.042310}, the authors have investigated the entanglement of mixed states for  quantum information purposes. However, in the context of quantum optics, study of entanglements arising between mixed states of radiation is not that much addressed which it deserves.

Sivakumar compared the Glauber-Lachs superposition with the mixed thermal coherent state (MTCS) at the level of density operator \cite{sivakumar2012effect}. The atomic inversion and the entanglement dynamics of both the states were compared, and it was reported that the MTCS is more sensitive to the thermal photon addition as opposed to the thermal photon addition in the G-L mixing. On the other hand in \cite{Rastegar_2016}, the authors have considered the atomic state to be the convex combinations of two density operators and investigated the atomic coherence with field in the Glauber-Lachs state.

Various studies have been carried out on the squeezed coherent states \cite{RevModPhys.58.1001, PhysRevA.47.5138, PhysRevA.61.010303, PhysRevA.61.022309, PhysRevLett.80.869, PhysRevA.60.937} i.e., here the PSCS. Mouloudakis and Lambropoulos have studied the squeezed coherent states in double optical resonance \cite{photonics8030072}; Li-Yun Hu and Zhi-Ming Zhang have investigated the statistical properties of photon-added two modes squeezed coherent states \cite{hu2013statistical}. Studies have been done to create entangled coherent states by mixing squeezed vacuum and coherent light \cite{israel2019entangled} and to find the time-evolution of squeezed coherent states of a generalized quantum parametric oscillator. Also, in quantum information and computation, the squeezed coherent states have been used \cite{PhysRevA.87.012307, PODOSHVEDOV2013192}. 

The motivation for this work arises from the question: to what extent squeezing  affects coherence in a coherent state versus mixing and also to investigate and compare their corresponding atom-field interactions. In this work, the field is taken to be in a mixed squeezed coherent state (MSCS), which has not been studied so far. The coherent photons are treated as signals and the squeezed photons as noise. In this work, the PCD, the atomic inversion and the entanglement dynamics for MSCS have been evaluated and these properties have been compared with those of the PSCS. Such a study is interesting as the results are quite contrary to those corresponding to the PSCS.

The organisation of this paper is as follows: in Section 2, MSCS is defined and the corresponding PCDs are obtained. In Section 3, the Jaynes-Cummings interaction of MSCS is studied. The focus is to bring out the differences in the evolution of the population inversion and the evolution of entanglement when the atom interacts with the PSCS and MSCS. In the following sections, the quadrature squeezing, Mandel's $Q$ parameter and the Wigner function of MSCS are also presented.

\section{Photon counting distribution (PCD)}
The  squeezed coherent states  are defined as
\begin{equation}
\ket{\alpha,\zeta}\equiv\hat{D}(\alpha)\hat{S}(\zeta)\ket0,
\end{equation}
where
\begin{equation}
\hat{D}(\alpha) = \exp(\alpha \hat{a}^{\dagger} - \alpha^{*} \hat{a}),
\end{equation}
is the displacement operator, for $\alpha$  a complex parameter; $\hat{a}$ and $\hat{a}^{\dagger}$ are the photon annihilation and creation operators respectively and
\begin{equation} 
\hat{S}(\zeta) = \exp\left(-\frac{1}{2}\zeta \hat{a}^{\dagger2} + \frac{1}{2} \zeta^{*}\hat{a}^{2}\right).
\end{equation}
The density matrices for pure squeezed coherent states (PSCS) and mixed squeezed coherent states (MSCS) are given by,
\begin{equation}
\hat{\rho}_{\text{pure}} =  \ket{\alpha,\zeta}\bra{\alpha,\zeta},
\end{equation}
and 
\begin{equation}
\hat{\rho}_{\text{mixed}} =  q\ket{\alpha}\bra{\alpha} + (1-q)\ket\zeta\bra\zeta, 
\end{equation}
where $q$ is the probability of the field being in the coherent state and $(1-q)$ is the probability of being in the squeezed state.
In this study, the field is prepared in the MSCS as in Eq. (3) and various quantum optical quantities associated with these states are compared with the corresponding quantities of the PSCS.

To study the interaction dynamics of the atom with two different states of fields, we need to fix a parameter which is common to both fields. Here, for PSCS and MSCS, the average number of coherent photons $N_{c}$ and average number of squeezed photons $N_{s}$ are common, but $q$ which occurs only in MSCS is variable. It is essential to fix the value of $q$ so that the states have the same mean number of photons. But, for PSCS $\langle n \rangle = N_c + N_s$ and for MSCS $\langle n \rangle = q N_c + (1-q) N_s$. So, by equating the mean number of photons we will never get a solution for $q$. Another way to choose the value of $q$ is by equating  the overlap with the coherent state for both the PSCS and MSCS, i.e., $\bra{\alpha}\hat{\rho}_{\text{pure}}\ket{\alpha} = \bra{\alpha}\hat{\rho}_{\text{mixed}}\ket{\alpha}$. The contributions of coherent state to PSCS and MSCS are given by the following equations respectively;

\begin{eqnarray}
\bra{\alpha}\hat{\rho}_{\text{pure}}\ket{\alpha} &= \sech{|\zeta|},
\end{eqnarray}
    
\begin{multline}
 \bra{\alpha}\hat{\rho}_{\text{mixed}}\ket{\alpha} = q + (1 -q ) \sech{|\zeta|}\\
 \times \exp{ \left[-2 \left(\frac{\alpha^2_R}{\exp(-2|\zeta|)+1}+ \frac{\alpha^2_I}{\exp(2|\zeta|)+1} \right)\right]},
\end{multline}
where $\alpha_R$ and $\alpha_I$ are the real and imaginary parts of $\alpha$ respectively. Solving these equations for the real part of $\alpha$ we get 

\begin{equation}\label{eu_eqn6}
    q = \frac{\sech{|\zeta|} \left[1- \exp\left(-\alpha^2_R(1 + \tanh{|\zeta|}\right)\right]}{1- \sech{|\zeta|}\,\exp\left(-\alpha^2_R(1 + \tanh{|\zeta|})\right)}.
\end{equation}
From eqn.\ref{eu_eqn6} we observe that if $|\zeta| = 0$, $q$ becomes $1$ which makes both PSCS and MSCS pure coherent state. But, if $\alpha = 0$ then $q=0$ and both the states become pure squeezed states. So, the equal overlap of the coherent state implies that the limiting cases of PSCS and MSCS are the same.

The PCD for the PSCS is given by ~\cite{subeesh2012effect, loudon1987squeezed, PhysRevA.13.2226},

\begin{equation}
P(n)=\frac{1}{n!\mu}\left(\frac{\nu}{2\mu}\right)^{2}H^{2}_{n}\left(\frac{\beta}{\sqrt{2\mu\nu}}\right) \exp\left(-\beta^{2}\left(1-\frac{\nu}{\mu}\right)\right),
\end{equation}
where $\mu=\cosh|\zeta|=\sqrt{1+N_{s}}$, $\nu=\sinh|\zeta|=\sqrt{N_{s}}$ and $\beta=\sqrt{N_{c}}(\sqrt{1+N_{s}}+\sqrt{N_{s}})$; $N_{s}=$ average number of squeezed photons and $N_{c}=$ average number of coherent photons. 

The PCD for the MSCS is given by \cite{pujamscdissertation},
\begin{equation}
P(n)=q|\braket{n|\alpha}|^{2}+(1-q)|\braket{n|\zeta}|^{2},
\end{equation}

\begin{eqnarray}
P(n)=
     \begin{cases}
     q\exp(-N_c)\frac{N_{c}^{n}}{n!}+\frac{(1-q)}{\sqrt{1+N_s}}\frac{n!}{2^{n}(\frac{n}{2}!)^{2}}\\
\times \left(\sqrt{\frac{N_s}{1+N_s}}\right)^{n}, \hspace{0.5cm}\text{for n is even}\\
q\exp(-N_c)\frac{N_{c}^{n}}{n!}, \hspace{0.5cm} \text{for n is odd}.
    \end{cases}       
    \end{eqnarray}
The average number of photons in MSCS is given by
\begin{equation}
\bar{n}_{avg}= qN_{c}+(1-q)N_{s} .
\end{equation}

\begin{figure}[h!]
\centering
\includegraphics[scale = 0.5]{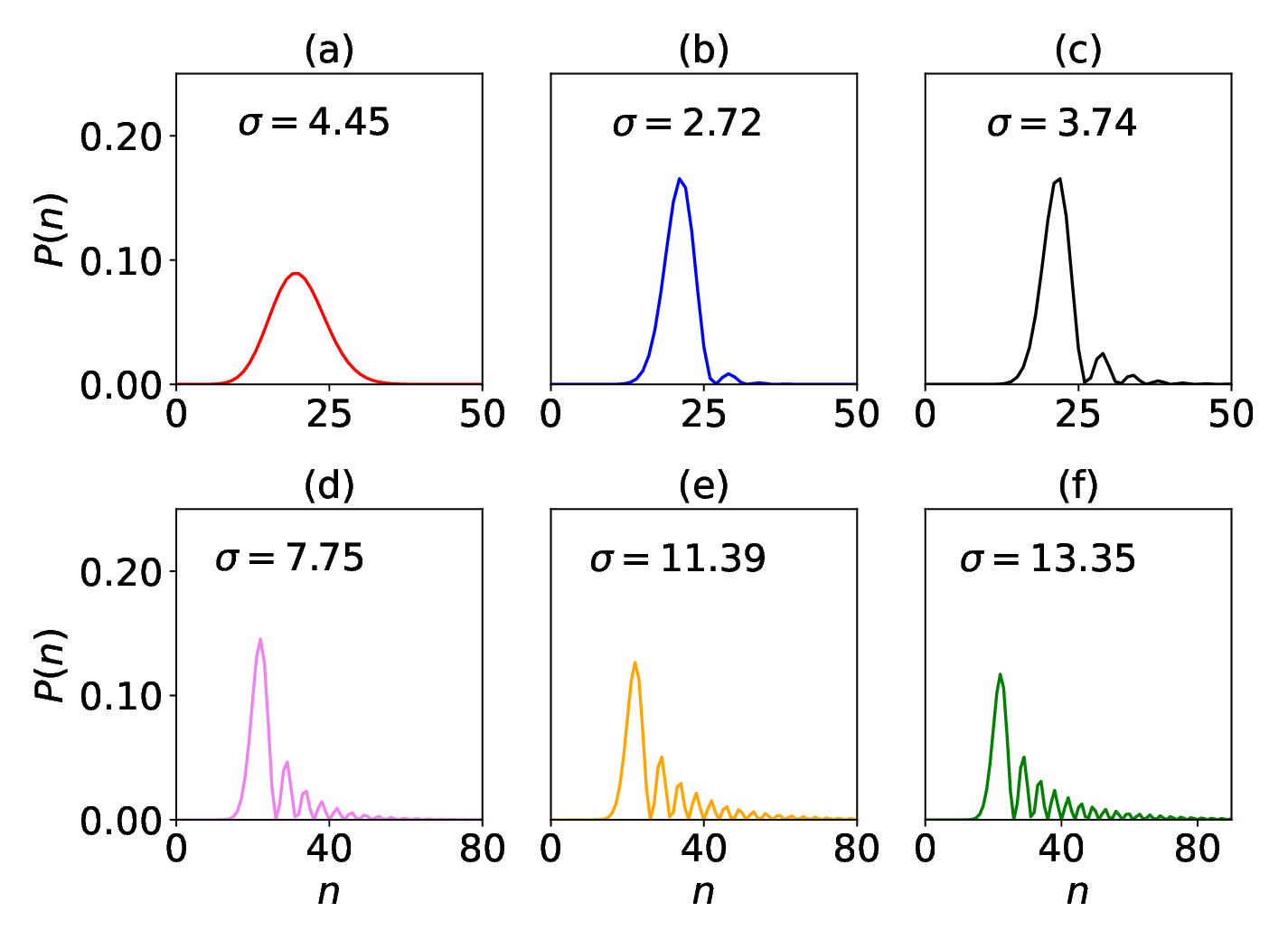}
\caption*{\textbf{Fig. 1} Photon counting distribution of PSCS for $N_{c} = 20$, $N_{s} = 0, 1, 2, 5, 8, 10$.}
\end{figure}

\begin{figure}[h!]
\centering
\includegraphics[scale = 0.5]{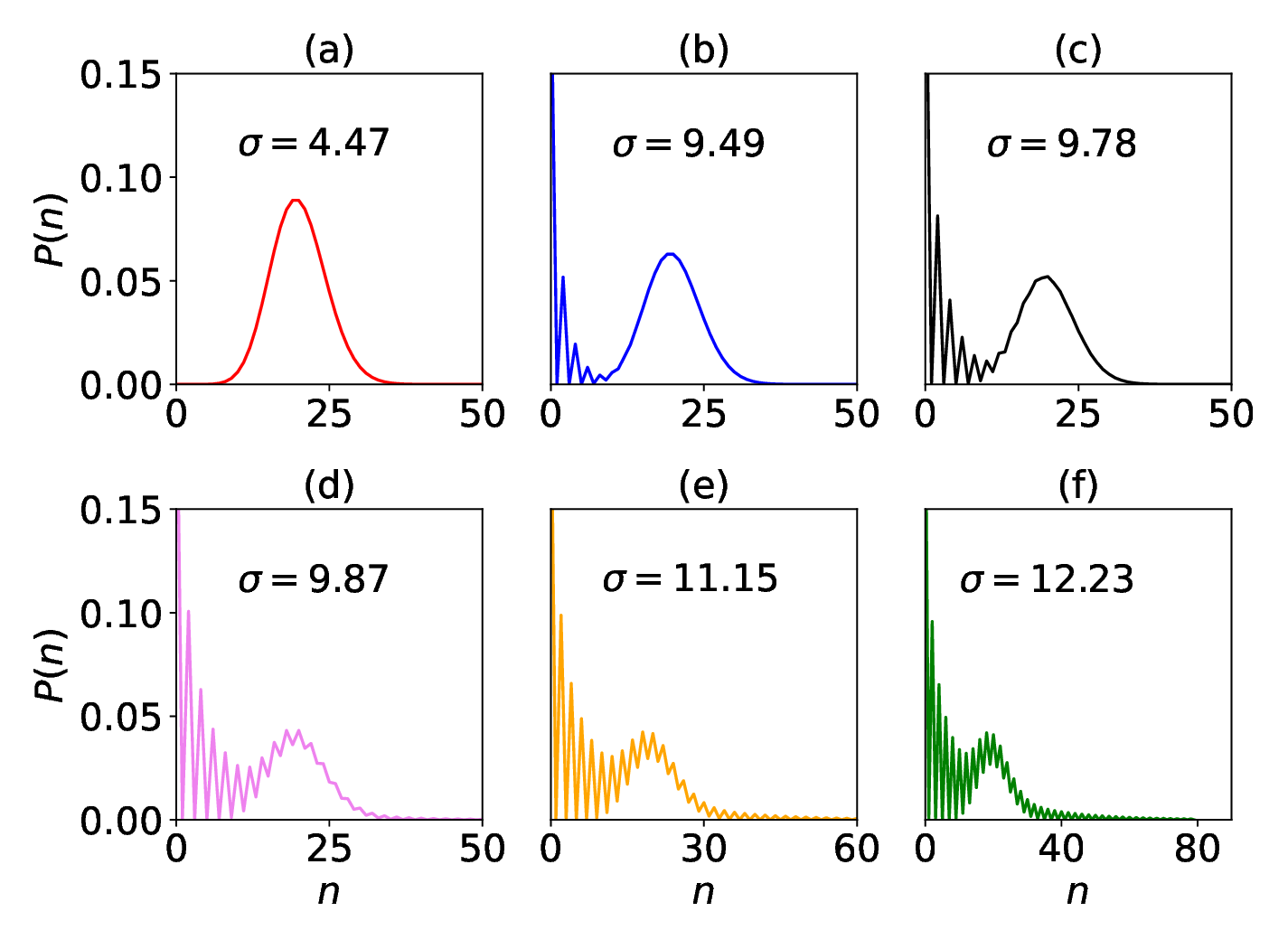}
\caption*{\textbf{Fig. 2}  Photon counting distribution of MSCS for $N_{c}=20$, $N_{s}=0,1,2,5,8,10$ and $q = 1.00, 0.70, 0.58, 0.41, 0.33, 0.30$}
\end{figure}

In Fig. 1, the PCD for PSCS is plotted for $N_{c} = 20$; $N_{s} = 0, 1, 2, 5, 8, 10$. It is observed from Fig. 1(a) that for $N_{s}=0$, it is just the $P(n)$ of a coherent state with $N_{c}=20$. Now, as $N_{s}$ is increased $P(n)$ changes significantly. In Fig. 1(b) it is observed that for $N_{s}=1$, the peak of the $P(n)$ becomes almost double the peak for $N_{s}=0$, i.e., the PCD gets localized. Also, it can be noted that the PCD begins to oscillate ~\cite{satyanarayana1989ringing}. If $N_{s}$ is increased further, the peak of $P(n)$ comes down and the oscillatory behaviour becomes very significant which is shown in Figs. 1(c), 1(d), 1(e) and 1(f). This oscillatory behaviour of the PCD is manifested in the atomic inversion and in the entanglement dynamics ~\cite{subeesh2012effect}. It is also noticeable that with increasing $N_{s}$, the peak of $P(n)$ also comes down.

Figure 2 represents the PCD for MSCS. For MSCS, in Fig. 2(a), $N_{c}=20$ and $N_{s}=0$ is chosen initially, i.e., $P(n)$ represents a coherent state distribution. If $N_{s}$ is increased to $ N_{s}=1$, it is observed from Fig. 2(b) that oscillations start at the beginning of the distribution, but there are no oscillations at the tail of the PCD. These oscillations appear from the distribution of the squeezed photons in the mixed states. In Fig. 2(c), $N_{c} = 20$ and $N_{s} = 2$; here it is seen that the amplitude of the oscillations increases and more part of the distributions take part in oscillation. If $N_{s}$ is increased further to $N_{s}=5$, the oscillatory behaviour - though not very pronounced - is seen till the tail of the PCD which is represented in Fig. 2(d). The addition of squeezed photons does not destroy the signature peak of the coherent state PCD. Instead, they introduce high-frequency oscillations enveloping the peak representing a coherent state. From Figs. 2(e) and 2(f) it is observed that on increasing the number of squeezed photons to $N_{s} = 8, 10$, the enveloping oscillations become more pronounced with a gradual decrease in the coherent state peak. The oscillations at the tail increase with $N_{s}$.

The main contrast in the behaviour of the $P(n)$ of MSCS is that  with the increase in squeezing, it set about with oscillations for small values of $n$. This behaviour is to be compared with  the tail of $P(n)$ of PSCS, which picks up oscillations, with the increase in squeezing. So, a very small amount of squeezing is dominant to bring in the oscillations in the $P(n)$ of MSCS, the tail picks up oscillations as in the case of PSCS. Another important fact that differentiates between the PCD of PSCS and the PCD of MSCS is that there is no localization of the PCD in the case of MSCS.

\section{Density matrix approach to the atom-field interaction}

The Jaynes-Cummings interaction Hamiltonian for the atom-field interaction is well studied in quantum optics  and it is given by ~\cite{jaynes1963comparison}

\begin{equation}
\hat{H}=\hbar {\omega} \hat{a}^{\dagger}\hat{a}+\frac{\hbar\omega_{0}}{2}\hat{\sigma}_{z}+\hbar \lambda(\hat{\sigma}_{+}\hat{a}+\hat{\sigma}_{-}\hat{a}^{\dagger}),
\end{equation}
where $ \hat{\sigma}_{+} $ ,  $ \hat{\sigma}_{-} $  and  $ \hat{\sigma}_{z} $ are the Pauli pseudospin operators; $\hat{a}$ and $\hat{a}^\dagger$ are the photon
annihilation and the photon creation operators; $\lambda$ is the coupling constant describing the atom-field interaction; $\omega$ is the field frequency and $\omega_{0}$ is the atomic transition frequency.

Under the interaction picture, if we use the resonant condition, i.e., the detuning $\Delta=\omega-\omega_{0}=0$, the interaction Hamiltonian becomes
\begin{equation}
\hat{H}_{I}=\hbar\lambda(\hat{\sigma}_{+}\hat{a}+\hat{\sigma}_{-}\hat{a}^{\dagger}).
\end{equation}
Let $\ket g$, $\ket e$ be the ground state and the excited state of the atom respectively and $\ket n$ are the Fock states of the radiation field. The action of $\hat{H}_{I}$ on the total initial state $\ket{e,n}$ of the atom-field system assuming that the atom initially in the excited state is given by the following equations,
\begin{eqnarray}
\hat{H}_{I}\ket{e,n}&=&\hbar\lambda\sqrt{n+1}\ket{g,n+1},\\
\hat{H}_{I}\ket{g,n+1}&=&\hbar\lambda\sqrt{n}\ket{e,n}.
\end{eqnarray}
We define $\hat{\rho}_{\text{tot}}(t)$ to be the total density operator of the atom-field system at time t, and the time evolution of this operator can be written as,
\begin{equation}
\hat{\rho}_{\text{\text{tot}}}(t)=\hat{U}(t)\hat{\rho}_{\text{tot}}(0)\hat{U}^{\dagger}(t),
\end{equation}
where $\hat{U}(t)=\exp(-\frac{\imath\hat{H}_{I}t}{\hbar})$ is the unitary time evolution operator.

$\hat{U}(t)$ can be expanded in the two dimensional subspace as \cite{gerry2005introductory}
\begin{equation}
\hat{U}(t)= \begin{pmatrix} \hat{C}(t) & \hat{S^{'}}(t)\\ \hat{S}(t) & \hat{C^{'}}(t) \end{pmatrix},
\end{equation}
where 
\begin{eqnarray}
\hat{C}(t) &=& \cos(\lambda t\sqrt{\hat{a}\hat{a}^\dagger}),\\
\hat{S}(t) &=& -\imath \frac{\hat{a}^{\dagger}\sin(\lambda t\sqrt{{\hat{a}\hat{a}^\dagger})}}{\sqrt{\hat{a}\hat{a}^\dagger}},\\
\hat{C^{'}}(t) &=& \cos(\lambda t\sqrt{\hat{a}^{\dagger}\hat{a}}),\\
\hat{S^{'}}(t) &=& -\imath \frac{\hat{a}\sin(\lambda t\sqrt{{\hat{a}^{\dagger}\hat{a}})}}{\sqrt{\hat{a}^{\dagger}\hat{a}}}.\\
\end{eqnarray}

Now, if $\hat{\rho}_{F}(0)$ is the density matrix for the field and $\hat{\rho}_{\text{atom}}$ is the density operator of the atom, then the initial density operator for the atom-field system is given by,
\begin{equation}
\hat{\rho}_{\text{tot}}(0)= \hat{\rho}_{F}(0)\otimes\hat{\rho}_{\text{atom}}.
\end{equation}
Initially, this atom-field system may be unentangled; but during time evolution the system may get entangled, which is a characteristic feature of the bipartite nature of the system. We assume that the atom is initially in the excited state $\ket e$, so $\hat{\rho}_{\text{atom}}(0)= \ket{e} \bra e$. In matrix form
\begin{equation}
\hat{\rho}_{\text{atom}}(0)=\begin{pmatrix} 1 & 0 \\ 0 & 0 \end{pmatrix}.
\end{equation}
Under the time evolution $\hat{\rho}_{\text{tot}}(t)$ becomes ~\cite{sivakumar2012effect, gerry2005introductory}
\begin{equation}
\hat{\rho}_{\text{tot}}(t)= \begin{pmatrix} \hat{C}(t)\hat{\rho}_{F}(0)\hat{C^{'}}(t) & -\hat{C}(t)\hat{\rho}_{F}(0)\hat{S^{'}}(t) \\ \hat{S}(t)\hat{\rho}_{F}(0) \hat{C}(t) & - \hat{S}(t)\hat{\rho}_{F}(0)\hat{S^{'}}(t) \end{pmatrix}.
\end{equation}

\section{Atomic inversion}
Ghoshal \textit{et. al,}\cite{PhysRevA.101.053805} found that a suppression of the collapse and revival of population inversion occurs in response to the insertion of Gaussian quenched disorder in atom-cavity interaction strength in the Jaynes-Cummings model. In\cite{WANG2018180}, the atomic inversion for squeezed vacuum state is studied in the Two-photon Jaynes-Cummings model. Obada \textit{et. al,}\cite{obada2018influence} studied the effect of intrinsic damping and classical field terms on atomic inversion. In\cite{ali2016some}, Ali \textit{et. al,} investigated the population inversion with the field in the finite-dimensional pair coherent state in the interaction of two two-level atoms with four-mode radiation filed interactions. In a very recent work Mandal \textit{et. al,} \cite{mandal2023atomic} atomic inversion and entanglement dynamics for squeezed coherent thermal states in the Jaynes-Cummings Model.

To calculate the atomic inversion $W(t)$, first we need to find the atomic density matrix $\hat{\rho}_{\text{atom}}$ from $\hat{\rho}_{\text{tot}}(t)$. This is done by tracing over the density matrix over the field state. So, 
\begin{eqnarray}
\hat{\rho}_{\text {atom}}(t)&=&Tr_{\text{field}}[\hat{\rho}_{\text{tot}}(t)]\\
&=& \sum_{n=0}^{\infty} \bra{n}\hat{\rho}_{\text{tot}}(t)\ket{n}.
\end{eqnarray}
The atomic inversion, which is defined as the difference in the probabilities of finding the atom in the excited state and ground state, is given as ~\cite{gerry2005introductory}
\begin{eqnarray}
W(t) &=& \langle \hat{\sigma}_{3}\rangle\\
 &=& Tr[\hat{\rho}_{\text{atom}}(t)\hat{\sigma}_{3}]\\
&=& \sum_{n=0}^{\infty} \bra{n}\hat{\rho}_{F}(0)\ket{n}\cos(2\lambda\sqrt{n+1}~t)\\
&=& \sum_{n=0}^{\infty}P(n)\cos(2\lambda\sqrt{n+1}~t).
\end{eqnarray}
So, the $W(t)$ for the PSCS is given by
\begin{equation}
W(t) = \sum_{n=0}^{\infty}\braket{n|\alpha,\zeta}\braket{\alpha,\zeta|n} \cos(2\lambda\sqrt{n+1}~t),
\end{equation}
where $|\braket{n|\alpha,\zeta}|^{2}$ is the PCD for PSCS given by Eq.(7).
For the MSCS, the atomic inversion $W(t)$ is given by

\begin{align}
W(t) =& \sum_{n=0}^{\infty} \left\{q\exp(-N_c)\frac{N_c^{2n}}{(2n)!}+\frac{(1-q)}{\sqrt{1+N_s}}\frac{(2n)!}{2^{2n}(n!)^{2}}\left(\frac{N_s}{1+N_s}\right)^{n}\right\}\nonumber\\
&\times \cos(2\lambda t\sqrt{2n+1})+
\left\{q\exp(-N_c)\frac{N_c^{2n+1}}{(2n+1)!}\right\}
\cos(2\lambda t\sqrt{2n+2}),
\end{align}
where $N_{c}$ is the average number of coherent photons and  $ N_s $ is average number of squeezed photons.

\begin{figure}[h!]
\centering
\includegraphics[scale = 0.5]{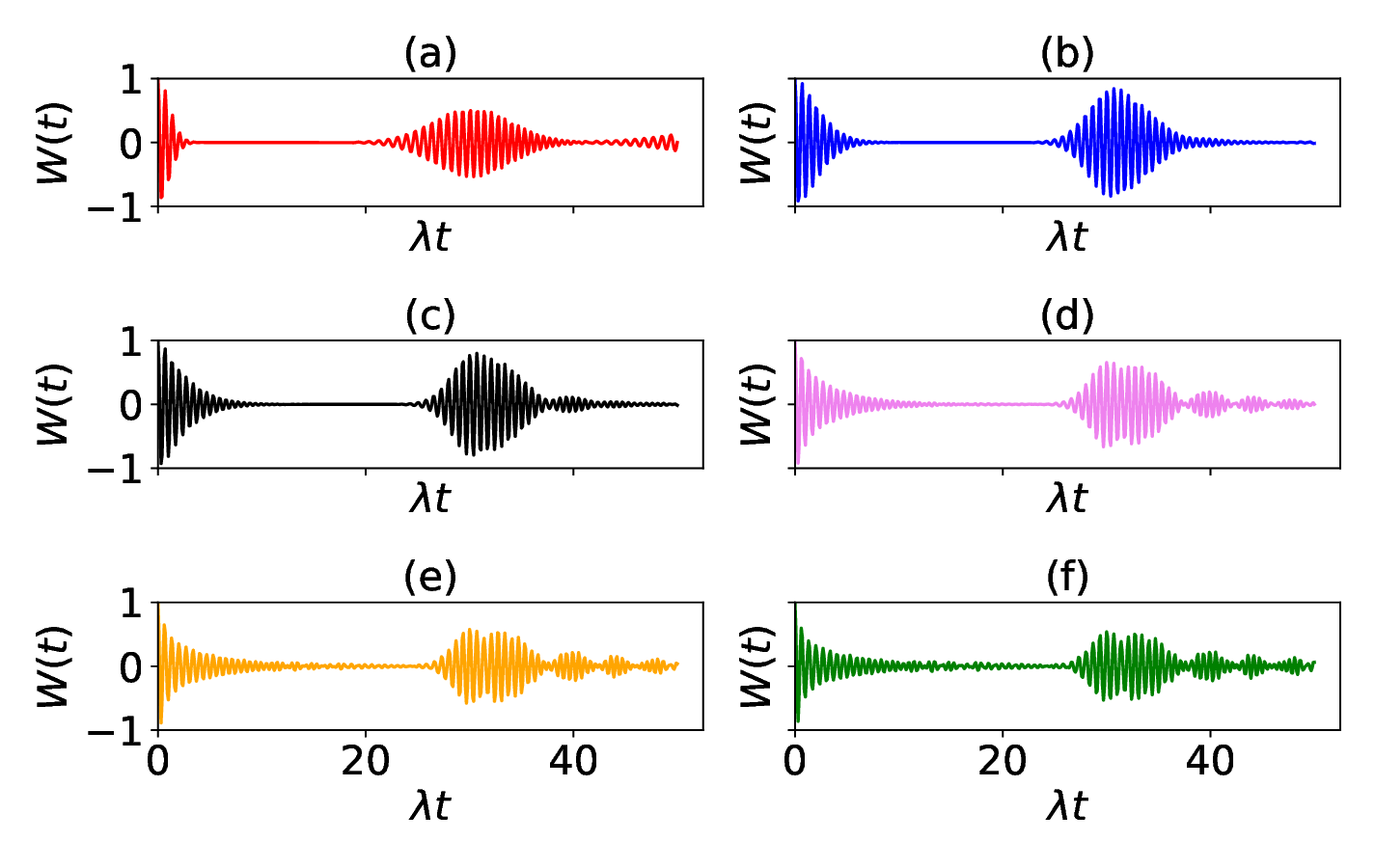}
\caption*{\textbf{Fig. 3} Atomic inversion $W(t)$ $vs$ $\lambda t$  for PSCS for $N_{c}=20$, $N_{s}=0,1,2,5, 8, 10$.}
\end{figure}

\begin{figure}[h!]
\centering
\includegraphics[scale = 0.5]{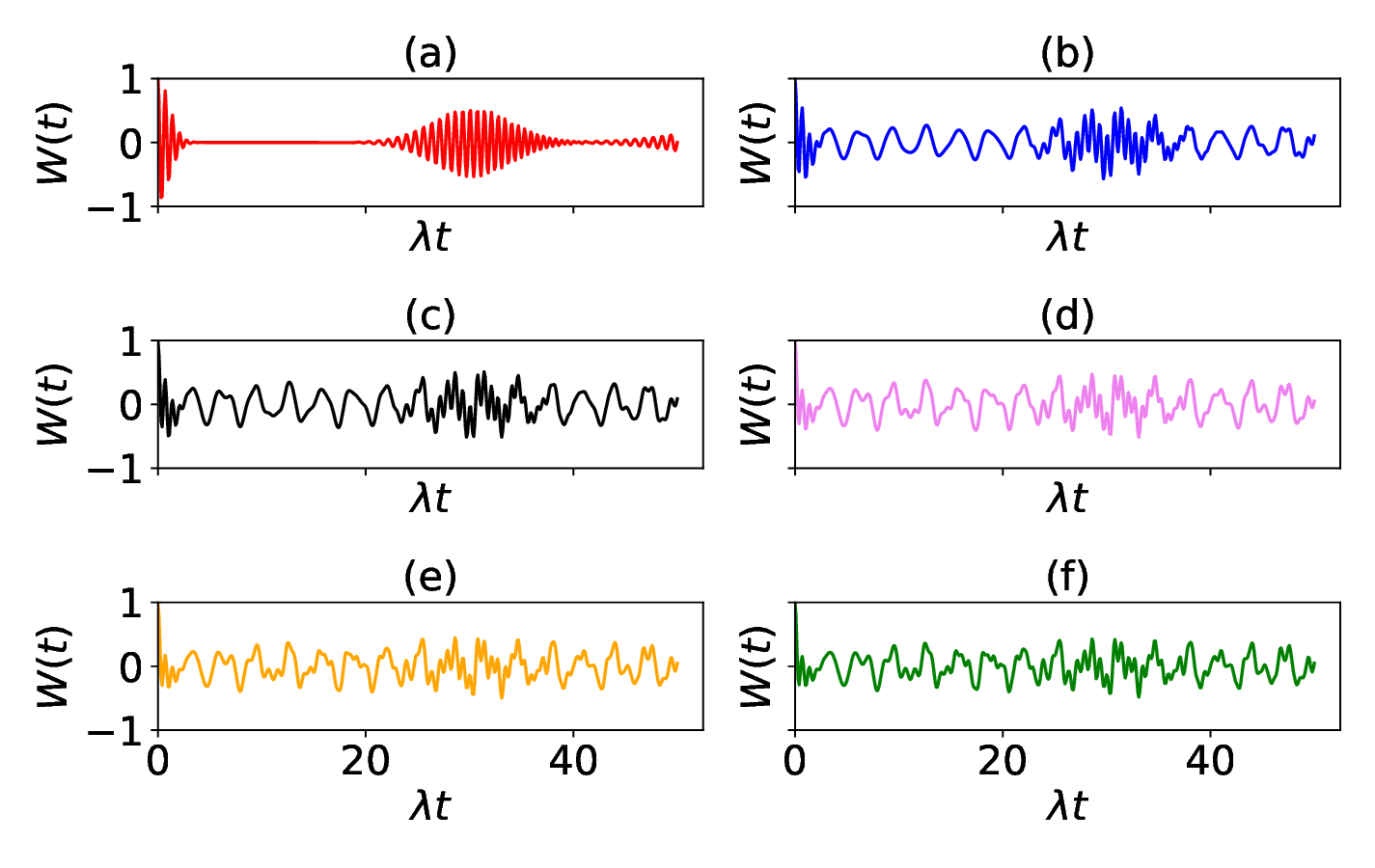}
\caption*{\textbf{Fig. 4} Atomic inversion $W(t)$ $vs$ $\lambda t$  for MSCS for $N_{c}=20$, $N_{s}=0,1,2,5, 8, 10$ and $q = 1.00, 0.70, 0.58, 0.41, 0.33, 0.30$.}
\end{figure}

The temporal variations of atomic inversion for both the PSCS and the MSCS are shown in Figs. 3 and 4 respectively. We have plotted  $W(t)$ $vs$ $\lambda t$, where $\lambda t$ is the scaled time. To investigate the  atomic inversion for both the states MSCS and PSCS, $N_{c}=20$ and $N_{s}=0,1,2,5, 8, 10$ are chosen which are the same numbers used in the case of the study of PCD. It can be observed from the Figs. 3(a) and 4(a) that $W(t)$ for both the states starts from that of a coherent state; but the addition of squeezing to the fields has very different behaviour compared to each other. In the case of PSCS, if a single squeezed photon is added to the PSCS, it can be seen from Fig. 4(b) that the collapse time of $W(t)$ becomes a little smaller with increasing oscillations. On increasing the squeezing to $N_{s} = 2, 5$, the collapse time gets smaller with ringing revivals on its dynamics which are represented in Figs. 3(c) and 3(d). If further squeezing is added to the system, from Figs. 3(e) and 3(f) where $N_{s} = 8, 10$, it is noticed that the whole collapse time becomes oscillatory with more ringing revivals in the dynamics.

In the case of MSCS, $W(t)$ behaves differently. From Fig. 4(b) it can be observed that if a single squeezed photon is added to the system, the collapse phenomena gets destroyed completely. If the squeezing is increased further, the pattern remains almost the same with no ringing revivals in its dynamics which can be seen in Figs. 4(c), 4(d), 4(e) and 4(f). From the behaviours of $W(t)$ for both the states, it can be concluded that the addition of squeezing has more sensitive effects on the atomic dynamics for MSCS as compared to the case of PSCS.

\section{Entanglement dynamics}
Investigating entanglement dynamics in quantum optical systems containing atoms and fields is one of the important aspects of quantum optics. Various kinds of studies are being done to measure entanglement between different subsystems in a quantum optical system. In\cite{liao2019properties, ALOTAIBI2022105540}, the entanglement between two-level atom and radiation field in an optomechanical system is investigated. Eberly \textit{et. al}\cite{eberly2006} showed the entanglement sudden death (ESD) in atom-atom entanglement in the double Jaynes-Cummings model. \cite{obada2018influence}, the phenomena of entanglement decay, sudden rebirth, and sudden death under the influence of intrinsic decoherence have been studied. In \cite{Pandit_2018, PhysRevA.101.053805}, the effects of photon exchange and Ising interaction on entanglement dynamics have been studied. Mandal \textit{et. al,} \cite{Mandal_2024} have studied the role of thermal and squeezed photons in the entanglement dynamics of the double Jaynes–Cummings model.  Different measures of entanglement are there depending on the system under consideration. To find the entanglement for the mixed states, Negativity is a suitable quantity. Negativity $N(t)$ is defined as the absolute sum of the negative eigenvalues of the partially transposed density operator $\hat{\rho}_{\text{tot}}^{\text{PT}}$ ~\cite{PhysRevA.67.022110, nielsen2002quantum}.
If, $\lambda_{k}$ are the eigenvalues of $\hat{\rho}_{\text{tot}}^{\text{PT}}$, then $N(t)$ is given by

\begin{equation}
N(t)=\sum_{k}\left[|\lambda_{k}|-\lambda_{k}\right]/2.
\end{equation}

\begin{figure}[h!]
\centering
\includegraphics[scale = 0.5]{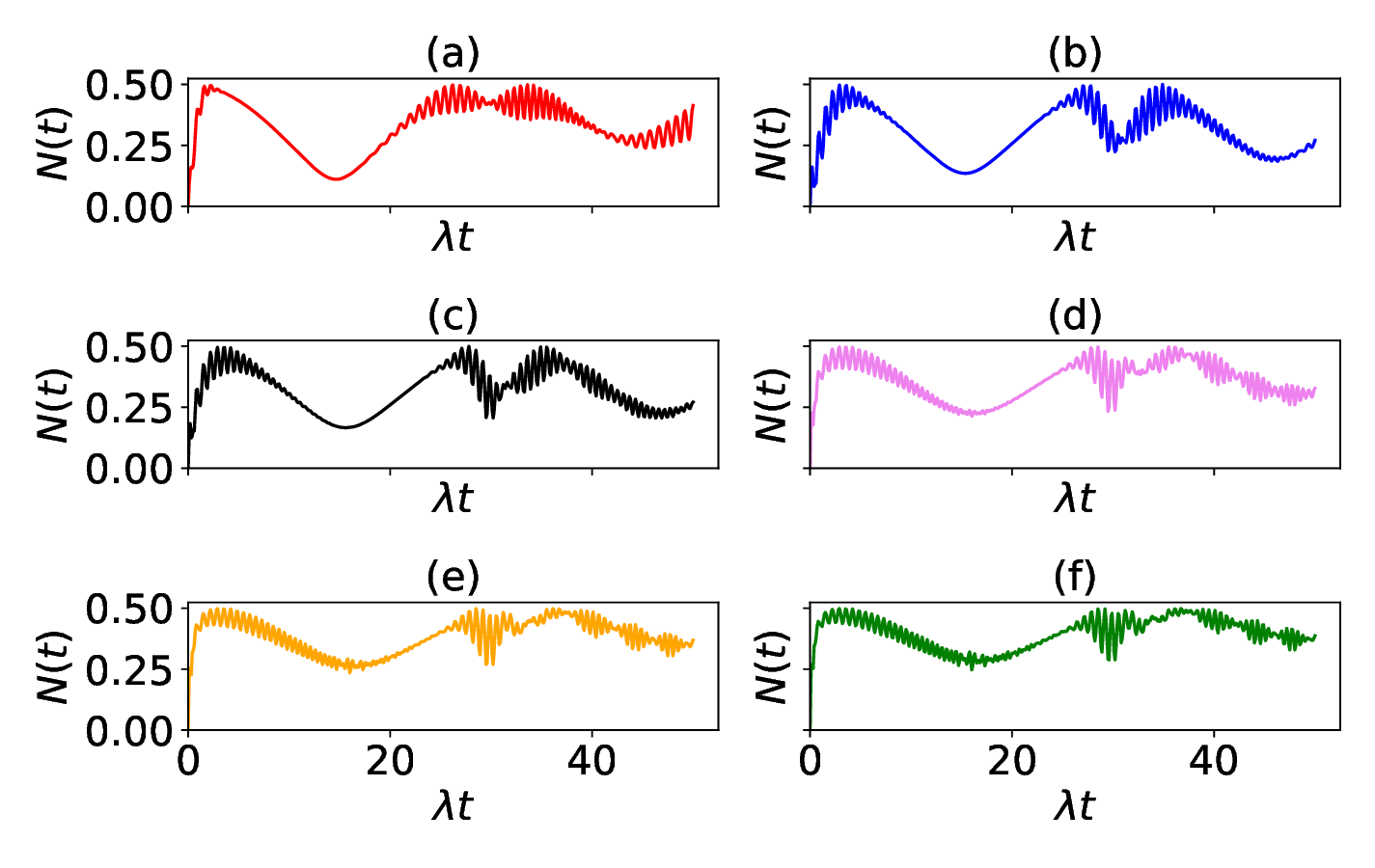}
\caption*{\textbf{Fig. 5} Entanglement dynamics $N(t)$ $vs$ $\lambda t$  for PSCS for $N_{c}=20$, $N_{s}=0,1,2,5, 8, 10$.}
\end{figure}

\begin{figure}[h!]
\centering
\includegraphics[scale = 0.5]{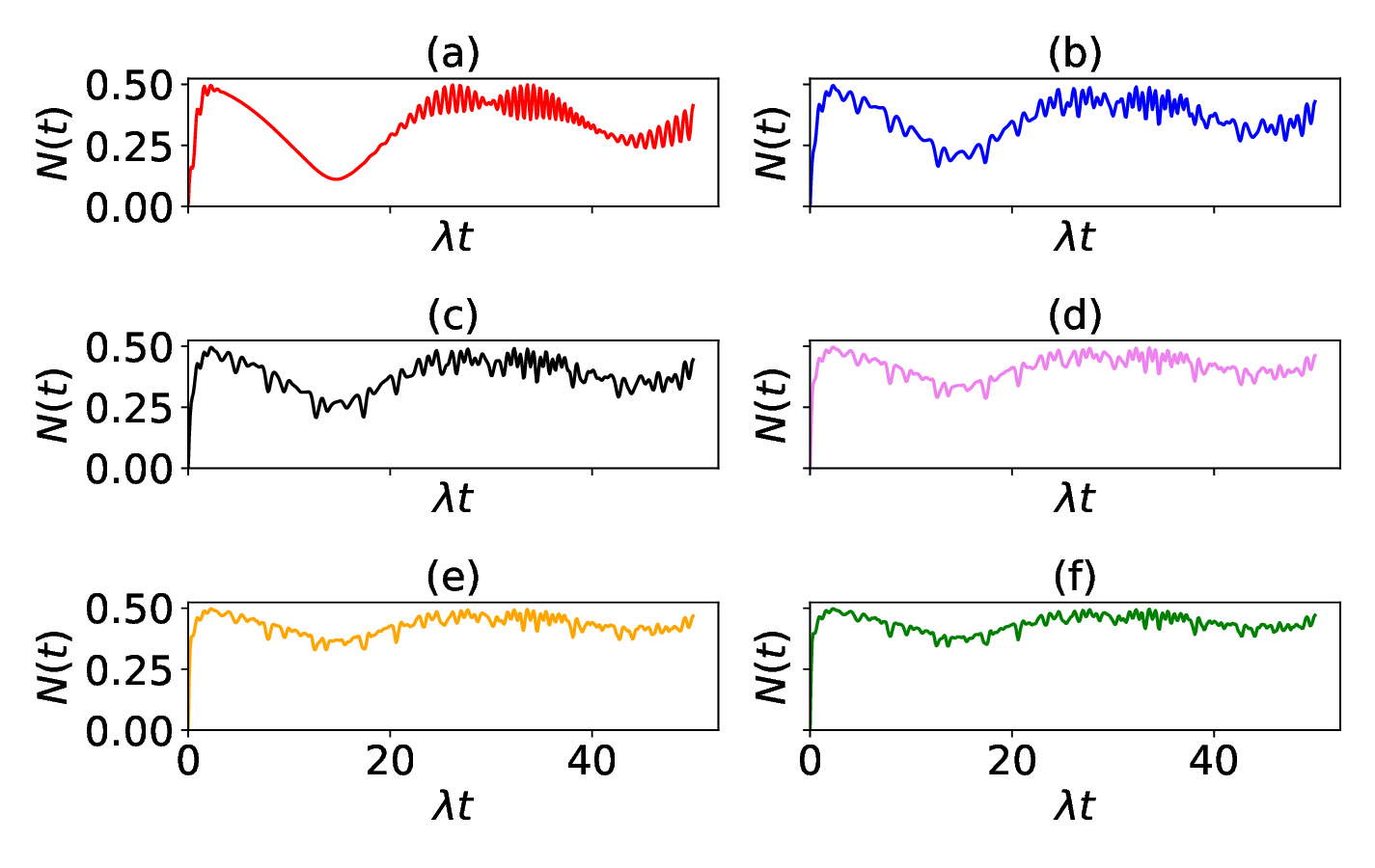}
\caption*{\textbf{Fig. 6} Entanglement dynamics $N(t)$ $vs$ $\lambda t$  for MSCS for $N_{c}=20$, $N_{s}=0,1,2,5, 8, 10$ and $q = 1.00, 0.70, 0.58, 0.41, 0.33, 0.30$.}
\end{figure}

Figures 5 and 6 represent the entanglement dynamics for the PSCS and the MSCS respectively. Like the atomic inversion, here also we have taken  the average number of coherent photons $N_{c}=20 $, the squeezed photons $N_{s}=0,1,2,5, 8, 10$ and $q = 1.00, 0.70, 0.58, 0.41, 0.33,\\
0.30$. From Fig. 5(a) it is observed that for PSCS, $N(t)$ starts from a coherent state dynamics as like $W(t)$ (see Fig. 3(a)). On addition of squeezed photons in the field, slight oscillations build up in the smooth collapse part of the dynamics and the second collapse point comes down significantly which is depicted in Fig. 5(b). From Figs. 5(c) and 5(d), it is evident that with further increase in the squeezing, the minimum value of the first collapse of $N(t)$ gets higher and the oscillations in the collapse part become more prominent. This pattern continues for other higher values of $N_s = 8, 10$; Figs. 5(e) and 5(f) show that with these higher values of $N_{s}$ the whole collapse becomes oscillatory.

For the MSCS, $N(t)$ starts from the coherent state dynamics also which is depicted in Fig. 6(a). In this case, Fig. 6(b) shows that if a squeezed photon is added, all the collapse parts become oscillatory and the minimum value of $N(t)$ gets higher. On further increasing the squeezing, the amplitude of the oscillations in the dynamics becomes smaller and the minimum values become higher which can be observed from the Figs. 6(c), 6(d), 6(e) and 6(f). One interesting fact is that the amplitude of the oscillation of $N(t)$ for MSCS becomes smaller with increasing squeezing but for PSCS, the amplitude of oscillations in the dynamics of $N(t)$ becomes larger. Since, $N(t)$ represents the extent of entanglement of the atom with the radiation field, it is observed that as the squeezing is increased, this entanglement gets stronger even during the collapse part. This effect is significantly enhanced in MSCS as compared to PSCS. 

\section{Atomic Inversion and Entanglement Dynamics for a fixed value of weightage parameter($q$)}

In the last two sections we have investigated the atomic-inversion and the entanglement dynamics for different values of $q$ which is dependent on $N_c$ and $N_s$. In this section the effects of fixed value of $q$ on the atom-field interaction dynamics is studied. For this $q = 0.8$ is chosen, i.e., $80\%$ of the field is coherent which is justified if we consider the coherent states as the signal and squeezed states as noise.

\subsection{Photon Counting Distribution}

\begin{figure}[h!]
\centering
\includegraphics[scale = 0.5]{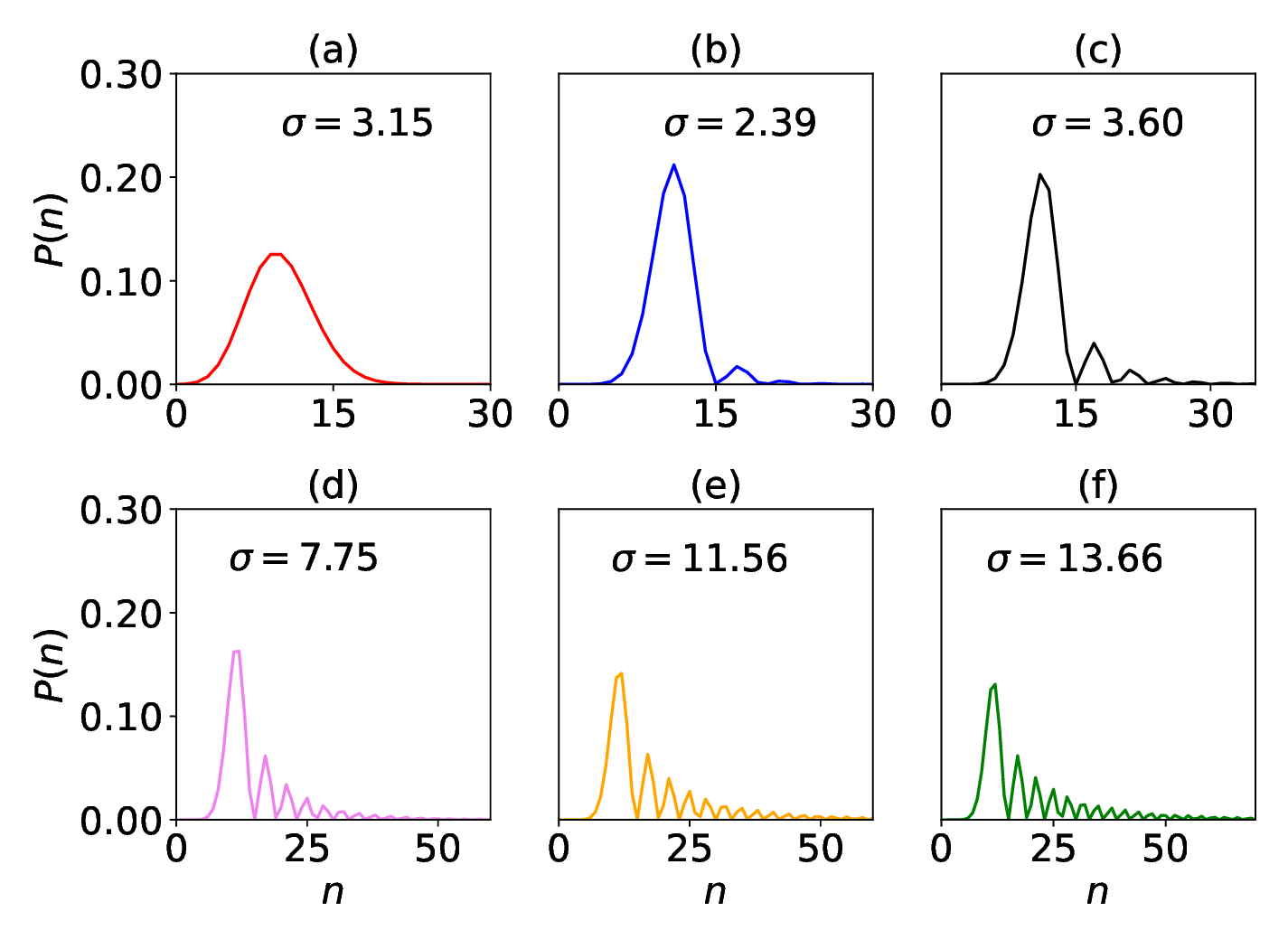}
\caption*{\textbf{Fig. 7} $P(n)$ $vs$ $n$  for PSCS for $N_{c}=10$, $N_{s}=0, 1, 2, 5, 8, 10$.}
\end{figure}

\begin{figure}[h!]
\centering
\includegraphics[scale = 0.5]{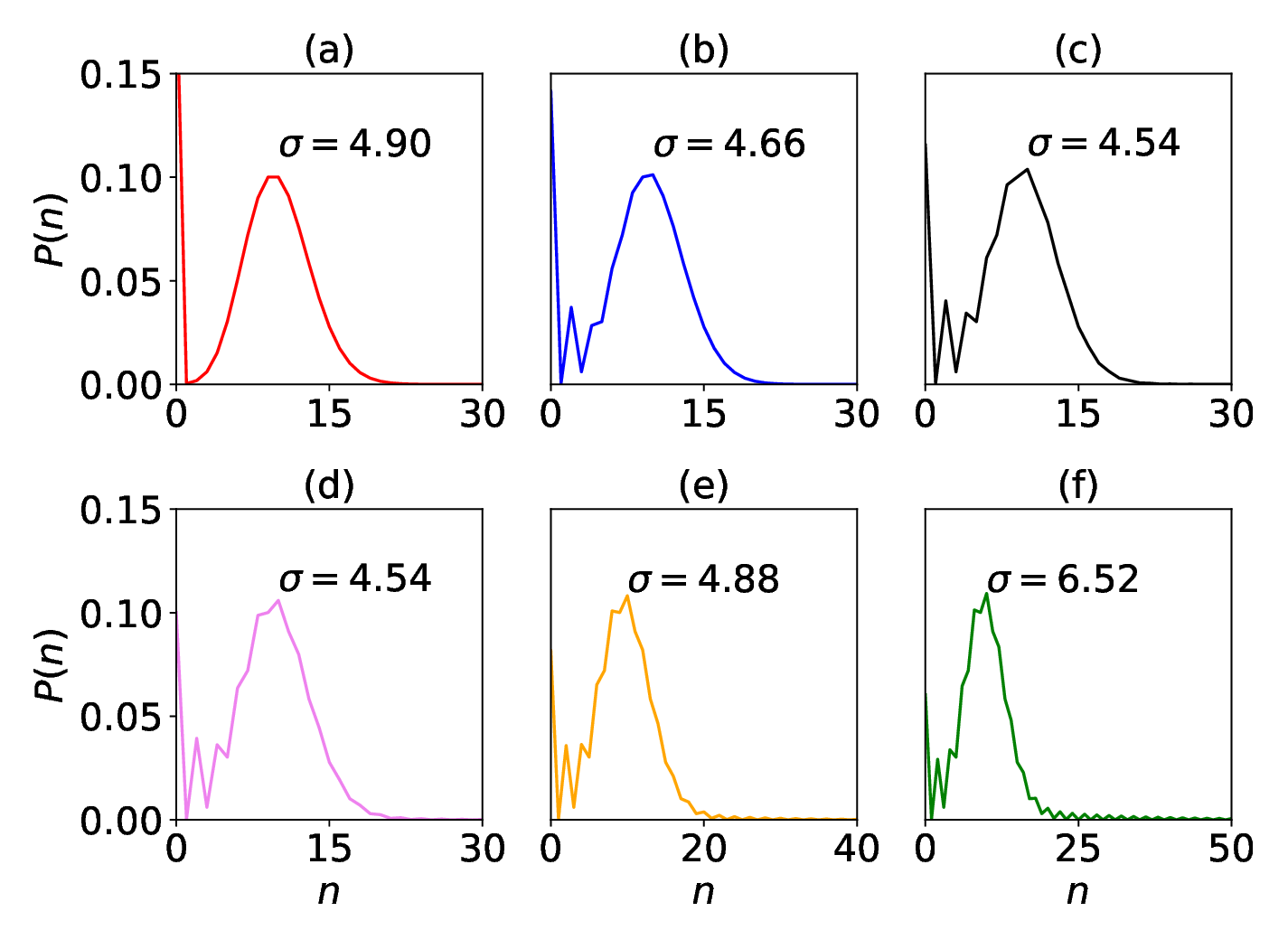}
\caption*{\textbf{Fig. 8}  $P(n)$ $vs$ $n$  for MSCS for $N_{c}=10$, $N_{s}=0, 1, 2, 5, 8, 10$ and $q = 0.8$.}
\end{figure}

The PCDs for PSCS and MSCS are plotted in the Figs. $7$ and $8$. For these plots $N_{c}=10, N_{s}=0,1,2,5, 8, 10$ and $q=0.8$ are chosen. It is observed that the PCD for PSCS for $N_{c} = 10$ mimics the pattern for $N_{c} = 20$; which are shown in Fig. 1 and Fig. 7; but for MSCS, the oscillations in the pcd are very small compared to the PCD for $N_{c} = 20$ (see Figs. 2 and 8). For MSCS, the PCD remains almost like a coherent state with little oscillations at the beginning and the tail of the PCD. 

\subsection{Atomic Inversion}
In order to investigate the atomic inversion for MSCS and PSCS, $N_{c}=10 $ and $N_{s}=0,1,2,5, 8, 10$ are taken. The atomic inversions for PSCS and MSCS for fixed $q$ are plotted in Figs. $9$ and $10$. As earlier, from Figs. 9(a) and 10(a), it is seen that here also $W(t)$ starts from the atomic dynamics of a coherent state for PSCS  but for MSCS it starts from an oscillatory pattern because this time the initial state is an  MSCS, a  mixture of coherent state and squeezed vacuum state. In the case of PSCS, with an increasing value of $N_{s}$ its pattern begins to get noisy with the larger amplitude of oscillations in the collapse part of the dynamics which are depicted in Figs. 9(b) and 9(c). It is also observed from Figs. 9(d), 9(e) and 9(f) that the duration of collapse time begins to decrease with increasing $N_{s}$ and ultimately goes away with a replaced ringing-revivals pattern. But, for MSCS, $W(t)$ shows exactly the opposite behaviour. If $N_{s}$ increases to $N_{s} = 2, 5$, Figs. 10(b) and 10(c) show that $W(t)$ for MSCS gradually develops a pattern with decreasing amplitude in the collapse part of the dynamics. Its pattern tends towards that of a coherent state to some extent for larger values of $N_{s}$ which is evident from Figs. 10(d), 10(e) and 10(f).

\begin{figure}[h!]
\centering
\includegraphics[scale = 0.5]{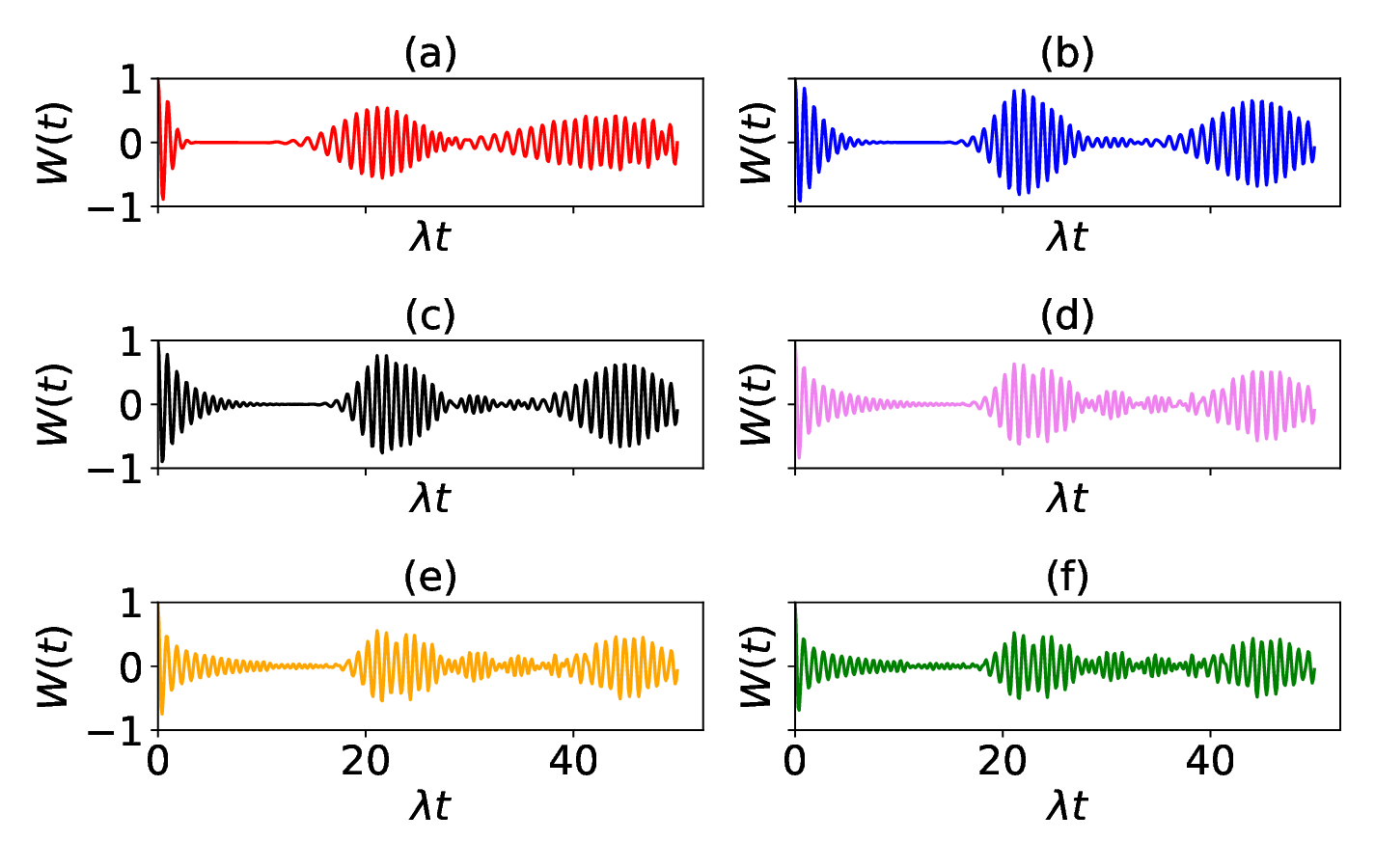}
\caption*{\textbf{Fig. 9} Atomic inversion $W(t)$ $vs$ $\lambda t$  for PSCS for $N_{c}=10$, $N_{s}=0,1,2,5, 8, 10$.}
\end{figure}

\begin{figure}[h!]
\centering
\includegraphics[scale = 0.5]{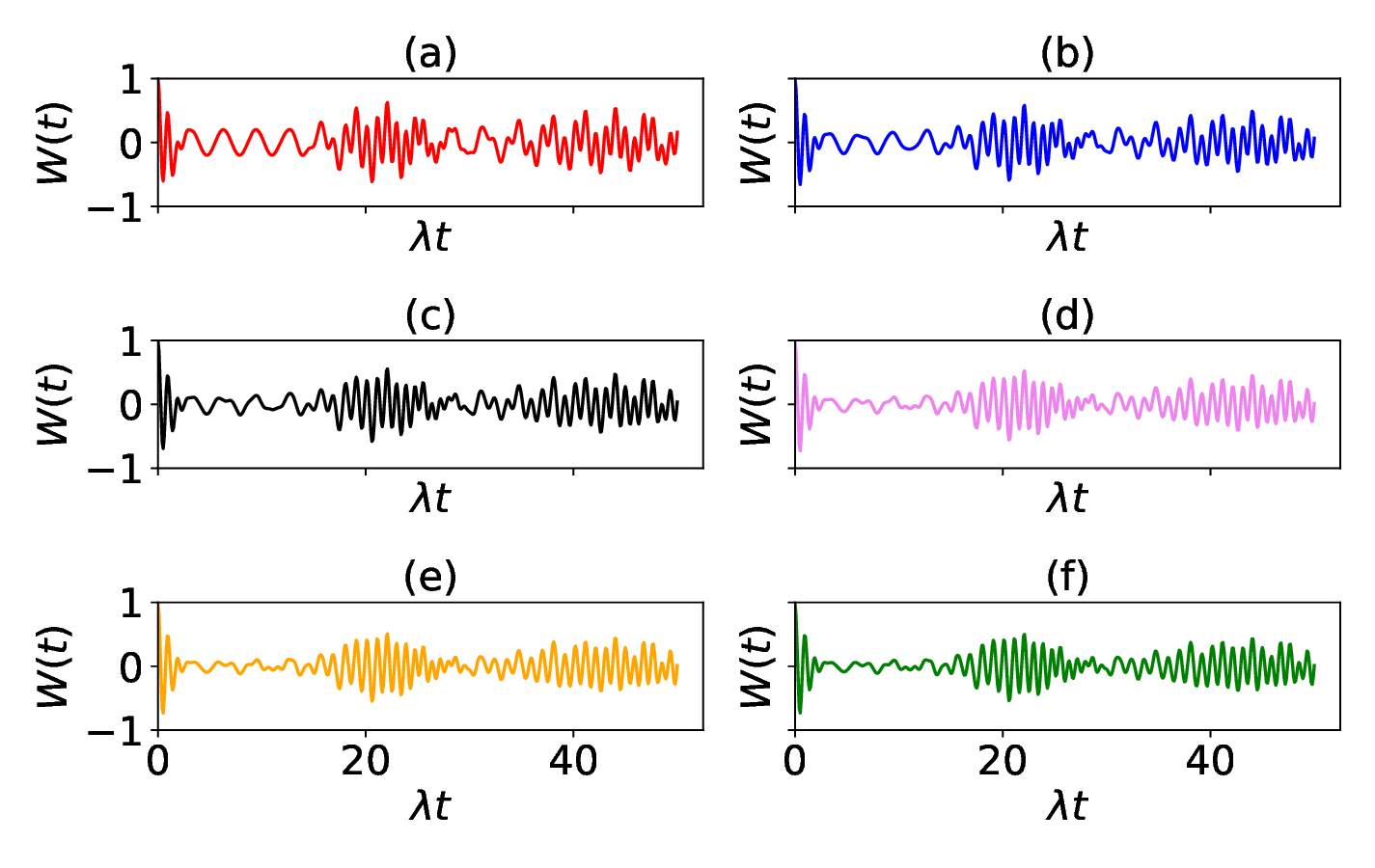}
\caption*{\textbf{Fig. 10} Atomic inversion $W(t)$ $vs$ $\lambda t$  for MSCS for $N_{c}=10$, $N_{s}=0,1,2,5, 8, 10$, $q = 0.8$.}
\end{figure}

\subsection{Entanglement dynamics}

\begin{figure}[h!]
\centering
\includegraphics[scale = 0.5]{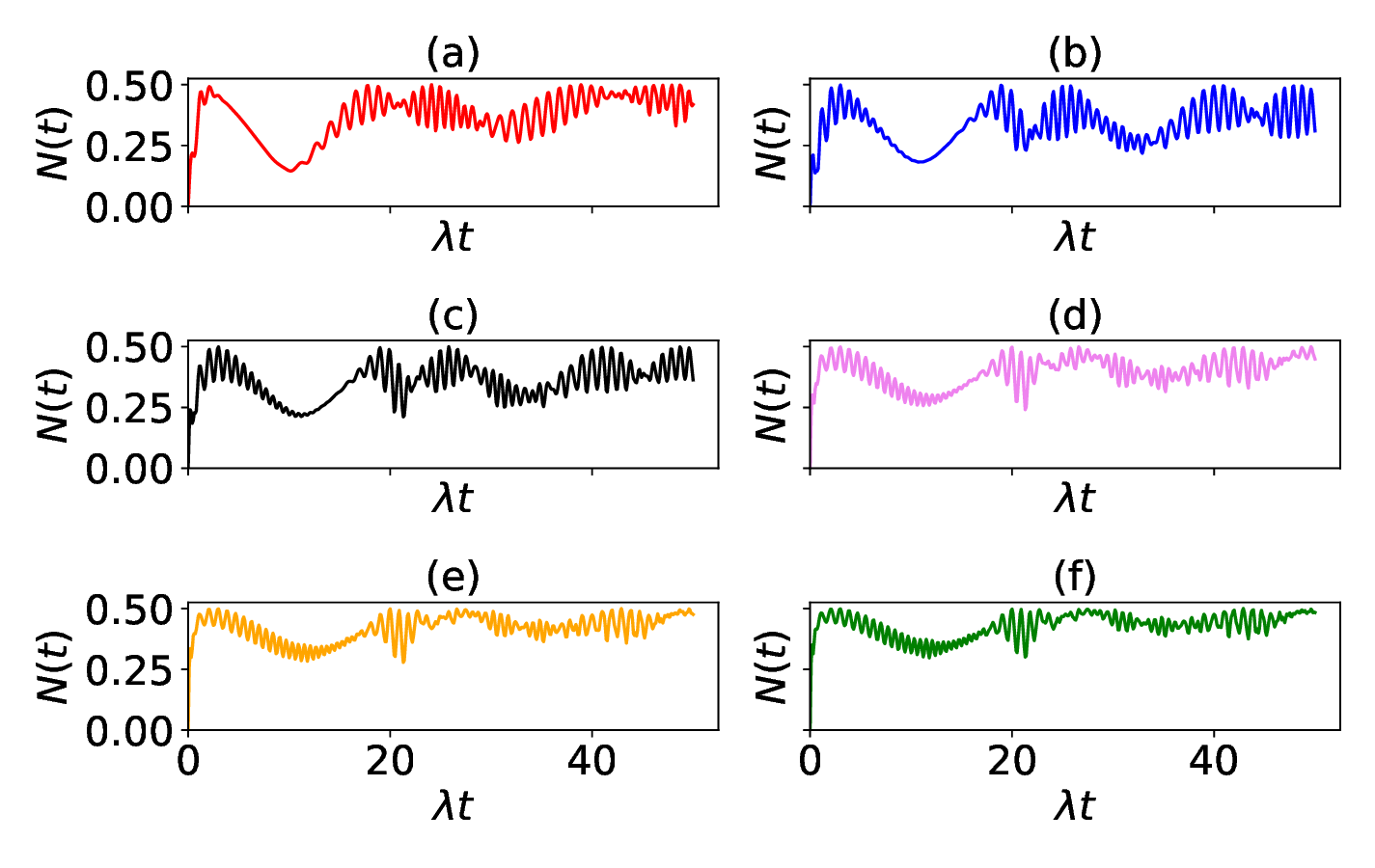}
\caption*{\textbf{Fig. 11} Entanglement dynamics $N(t)$ $vs$ $\lambda t$  for PSCS for $N_{c}=10$, $N_{s}=0,1,2,5, 8, 10$.}
\end{figure}

\begin{figure}[h!]
\centering
\includegraphics[scale = 0.5]{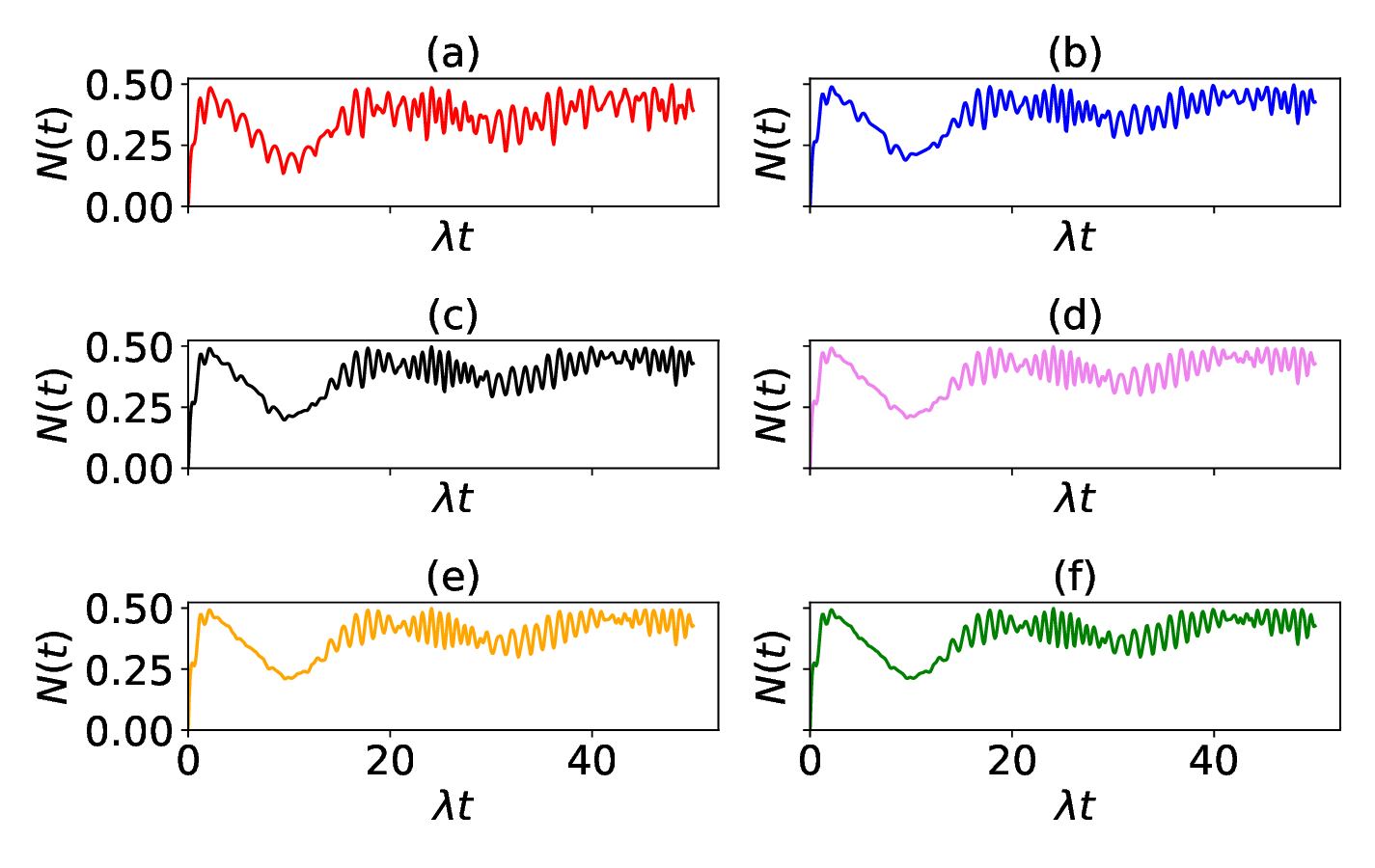}
\caption*{\textbf{Fig. 12} Entanglement dynamics $N(t)$ $vs$ $\lambda t$  for MSCS for $N_{c}=10$, $N_{s}=0,1,2,5, 8, 10$, $q=0.8$.}
\end{figure}

The temporal evolution of $N(t)$ $vs$ $\lambda t$ for PSCS and MSCS is shown in Figs. $11$ and $12$. Here, from Fig. 11(a) it is observed that for $N_{s} = 0$, $N(t)$ shows the coherent state dynamics for the PSCS, but, in Fig. 12(a), for MSCS, its behaviour is very noisy because of the presence of the squeezed vacuum state. As $N_{s}$ is increased, $N(t)$ behaves very differently for PSCS and MSCS. From the corresponding plots in Figs. 11(b) and 11(C), we see that for PSCS, $N(t)$ begins to  deviate from the coherent state dynamics and tends to a noisy behaviour. Ultimately, the whole dynamics becomes oscillatory for $N_{s} = 5, 8, 10$ as shown in Figs. 11(d), 11(e) and 11(f). Interestingly, for MSCS, it behaves exactly in the  opposite way. Here, $N(t)$ begins from a very noisy pattern and tends to the pattern of coherent state dynamics. It tends more rapidly to the coherent dynamics than the atomic inversion. It is very non-intuitive that it tends towards the classical state pattern due to the addition of quantum noise. 

\section{Mandel's $Q$ parameter}

\begin{figure}[h!]
\centering
\includegraphics[scale = 0.6]{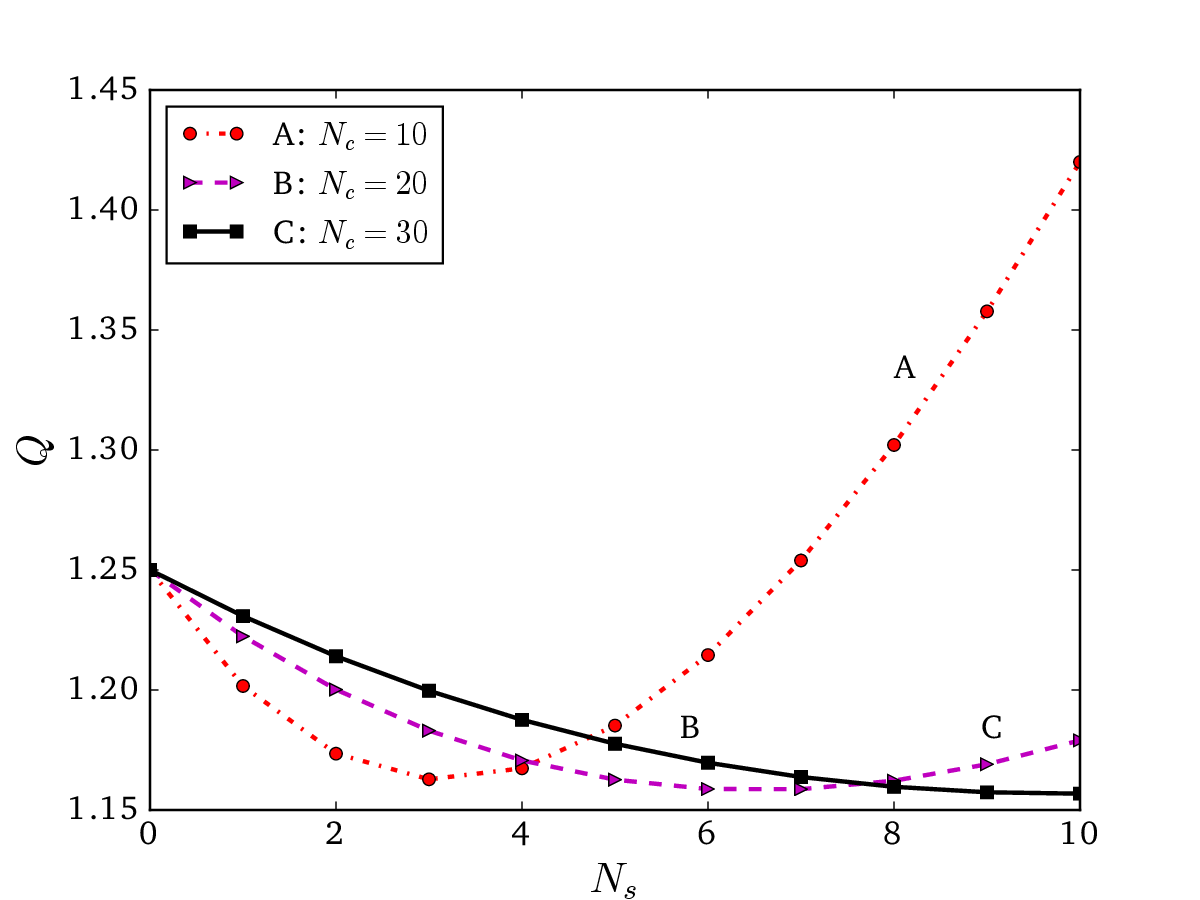}
\caption*{\textbf{Fig. 13} Mandel's $Q$ parameter $vs$ $N_{s}$ for MSCS for $N_{c}=10, 20,30$ and $q =0.8$.}
\end{figure}

\begin{figure}[h!]
\centering
\includegraphics[scale = 0.8]{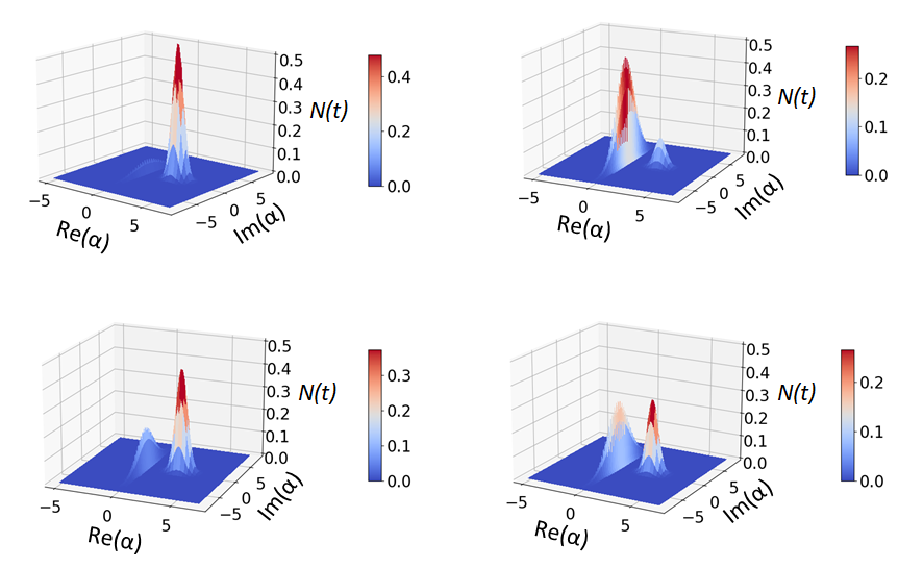}
\caption*{\textbf{Fig. 14} Wigner function for MSCS for $N_{c}=10$, $N_{s}=5$ and $q=0.2, 0.5, 0.7, 0.9$.}
\end{figure}
In this section, we investigate the Mandel's $Q$ parameter for MSCS. It is a measure of nonclassicality of the state of radiation field. Mandel's $Q$ parameter is defined by \cite{loudon, wolf, PhysRev.130.2529, Mandel:79}
\begin{equation}
Q=\frac{\braket{(\Delta n)^{2})}}{\braket{n}}-1,
\end{equation}
where $\braket n $ is the average number of photons. The plot of the variation of $Q$ with respect to $N_{s}$ for MSCS is shown in Fig. 13. The plot shows that $Q$ is always greater than zero. So, it can be concluded that MSCS shows super-Poissonian statistics. The variation in $Q$ is maximum for $N_{c}=10$ and minimum for $N_{c}=30$. For PSCS, $Q$ parameter has been calculated by Subeesh \textit{et al.} \cite{subeesh2012effect}. They observed that PSCS shows sub-Poissonian statistics around a particular value of $N_{s}$ for a specific value of $N_{c}$. This value of $N_{s}$ causes the localization in PCD for PSCS. But, for large values of $N_{s}$ it shows super-Poissonian statistics.

\section{Wigner function}

The Wigner function $W(\alpha)$ is well known in literature and it is defined as ~\cite{HILLERY1984121, LEE1995147, schleich, agarwal2013}
\begin{equation}
W(\alpha)=\frac{1}{\pi^{2}}\int d^{2}\beta\enspace \text{Tr}[\hat{\rho}\hat{D}(\beta)]\exp(\beta^{*}\alpha-\beta\alpha^{*}).
\end{equation}
The density operator for MSCS is given by
\begin{equation}
\hat{\rho}_{\text{mixed}} =  q\ket{\alpha}\bra{\alpha} + (1-q)\ket\zeta\bra\zeta.
\end{equation}
So, $W(\alpha)$ for this state is 
\begin{equation}
W(\alpha)=q\frac{2}{\pi}\exp\left(-2|\alpha-\gamma|^{2}\right)+(1-q)\frac{2}{\pi}\exp\left(-|\alpha\cosh |\zeta|-\alpha^{*}e^{\imath\phi}\sinh |\zeta||^{2}\right)
\end{equation}.

Figure 14 depicts $W(\alpha)$ for MSCS. It can be observed that the $W(\alpha)$ is a combination of two Gaussians at two different positions in phase space. One distribution is due to the coherent state and the other is due to the squeezed state. From these distributions, it can be noticed that $W(\alpha)$ is always positive at all points for MSCS. It is also well known that the Wigner distribution for PSCS is also positive.

\section{Conclusion}

We observe from the above study that various properties and the dynamics of atom-field interaction for  PSCS and  MSCS are very different and contrasting. The addition of the squeezed photons to the coherent photons has very different effects on the atomic inversion and entanglement dynamics for PSCS and MSCS. In the case of variable $q$, both the dynamics $W(t)$ and $N(t)$ starts from the dynamics of a coherent state field. But, with the addition of increasing value of squeezed photons, atomic-field dynamics for both the fields show oscillatory behaviour. It is noticed that for MSCS, the both dynamics are more sensitive to squeezed photons as compared to that of PSCS.

In the case of a fixed value of $q$, for PSCS, both the dynamics, i.e., the atomic inversion and the entanglement dynamics start from the pattern of a coherent state. On increasing the squeezed photons, the patterns gradually become noisy. Interestingly, for MSCS, it is quite the opposite. The PCD get appreciably delocalised In this case, the atomic inversion and the entanglement dynamics start from a very noisy behaviour, and tend towards that of a coherent state with increasing squeezed photons.

Preparation of convex combinations of density operators of spin-$1/2$ states is well known in lierature \cite{amitgoswami}. Experimental schemes to realise  MSCS is discussed in \cite{israel2019entangled}.

\section*{Acknowledgments}
We thank Dr. S. Sivakumar and Mr. Sushant Sharma for the discussions and suggestions.

\bibliographystyle{naturemag}
\bibliography{references_ijtp.bib}

\end{document}